\documentclass[prb,twocolumn,superscriptaddress,amsmath,amssymb, showkeys]{revtex4-2}

\usepackage{amssymb,amsmath}
\usepackage{color}
\usepackage{graphicx}%
\usepackage{dcolumn}%
\usepackage{bm}%
\usepackage{bbm}
\usepackage{latexsym,epsfig}
\usepackage{hyperref}
\usepackage{tabularx}
\usepackage{appendix}

\usepackage{marvosym}
\usepackage[caption=false]{subfig}

\usepackage{amsmath, amssymb, graphics, setspace}
\usepackage{graphicx}%
\usepackage{dcolumn}%
\usepackage{bm}%
\usepackage{lipsum}
\usepackage{tikz}
\usetikzlibrary{decorations.pathmorphing,decorations.markings, decorations.pathreplacing}
\usepackage{subcaption}

\newcommand{\z}{\textbf{z}}
\newcommand{\Uz}{\underline{\textbf{z}}}%
\newcommand{\U}[1]{\underline{#1}}
\newcommand{\x}{\textbf{x}}
\newcommand{\ps}{\mathrm{PS}}
\newcommand{\lps}{L\mathrm{PS}}
\newcommand{\Ulps}{\underline{L\mathrm{PS}}}
\newcommand{\hs}{\mathrm{HS}}
\newcommand{\Uhs}{\underline{\mathrm{HS}}}
\newcommand{\C}[1]{\overline{#1}}%

\begin{document}

\preprint{APS/123-QED}

\title{Cut-and-Project Density Functional Theory for Quasicrystals}

\author{Gavin N. Nop}
\affiliation{Department of Mathematics,  Iowa State University, Ames, Iowa 50011, USA}
\author{Jonathan D. H. Smith}
\affiliation{Department of Mathematics,  Iowa State University, Ames, Iowa 50011, USA}
\affiliation{The Ames National Laboratory, U.S. Department
of Energy, Iowa State University, Ames, IA 50011, USA}
\author{Thomas Koschny}
\affiliation{The Ames National Laboratory, U.S. Department
of Energy, Iowa State University, Ames, IA 50011, USA}
\affiliation{Department of Physics and Astronomy, Iowa State University, Ames, Iowa 50011, USA}
\author{Durga Paudyal}
\affiliation{Department of Physics and Astronomy, University of Iowa, Iowa City, Iowa 52242, USA}

\date{\today}%

\begin{abstract}
\noindent Cut-and-project from a symmetric structure in a higher-dimensional space is a standard method for describing the structure of a large class of quasicrystals. By means of a novel localization procedure, we now show how local physical interactions within these quasicrystals are also accurately described by cut-and-project, from corresponding physical interactions in the higher-dimensional space. 
A 
density functional theory (DFT++) formulation allows the cut-and-project method to handle the Schr\"odinger equation for interactions in quasicrystals. The theory is both rigorous and computationally tractable. The resulting \textit{ab initio} approach specifies  quasicrystalline quantum states, in contrast to previous approaches which only worked with crystalline approximants of the quasi-periodic structures.
\end{abstract}

\maketitle

\noindent 
A quasicrystal (QC) is characterized by a pure point diffraction pattern and lack of 3D translational symmetry
\cite{qc-discovery,history-2,QC-are-cool-1, paper-inspiration, topological-photonic-quasicrystals, antiferromagnetic-QC}. 
While the method of \emph{cut-and-project} (C+P)  from a notional crystalline structure in a higher-dimensional space is a standard technique for describing the structure of a large class of quasicrystals \cite{advanced-moire-experiment-bands, thue-morse, QCsymmetry, QCrystallographyGuide, firstAtomicStructure, aperiodic-diffraction, good-cut-and-project-characterization}, this approach has so far only proved feasible for analyzing their physics at the single-particle (or non-interacting particle, e.g. photonic) level \cite{important-tb-quasicrystal-photonics,paper-inspiration, history-7, soft-matter-numerics-use-for-dft, paper-preq-important, most-important-paper, history-6}.
Most notably, non-local terms in the Density Functional Theory (DFT) technique \cite{history-8} have frustrated naive attempts to connect C+P to DFT. Instead, quasicrystalline interactions have primarily been studied through the \emph{method of approximants}, using supercell simulations of crystalline approximants to the QC \cite{soft-matter-numerics-use-for-dft, paper-preq-important, characterization-1d-fractal-quasicrystal-spectrum}. These simulations have sometimes proved slow and difficult to bring to convergence \cite{QC-brute-force}.

In this work, we present three primary advances. First, we introduce the novel localization procedure that enables the cut-and-project method to handle physical theories which include interactions. The method reparametrizes differential equations in the higher dimensional space (HS), solves them by separation of the variables in the reparametrization, and then projects the solutions back down to the physical space (PS) in which the quasicrystal lies.

\begin{figure}[htbp]
    \begin{center}
        \begin{tabular}{@{}l l@{}}
            \small (a) &
            \raisebox{-0.95\height}{\includegraphics[width=0.35\textwidth]{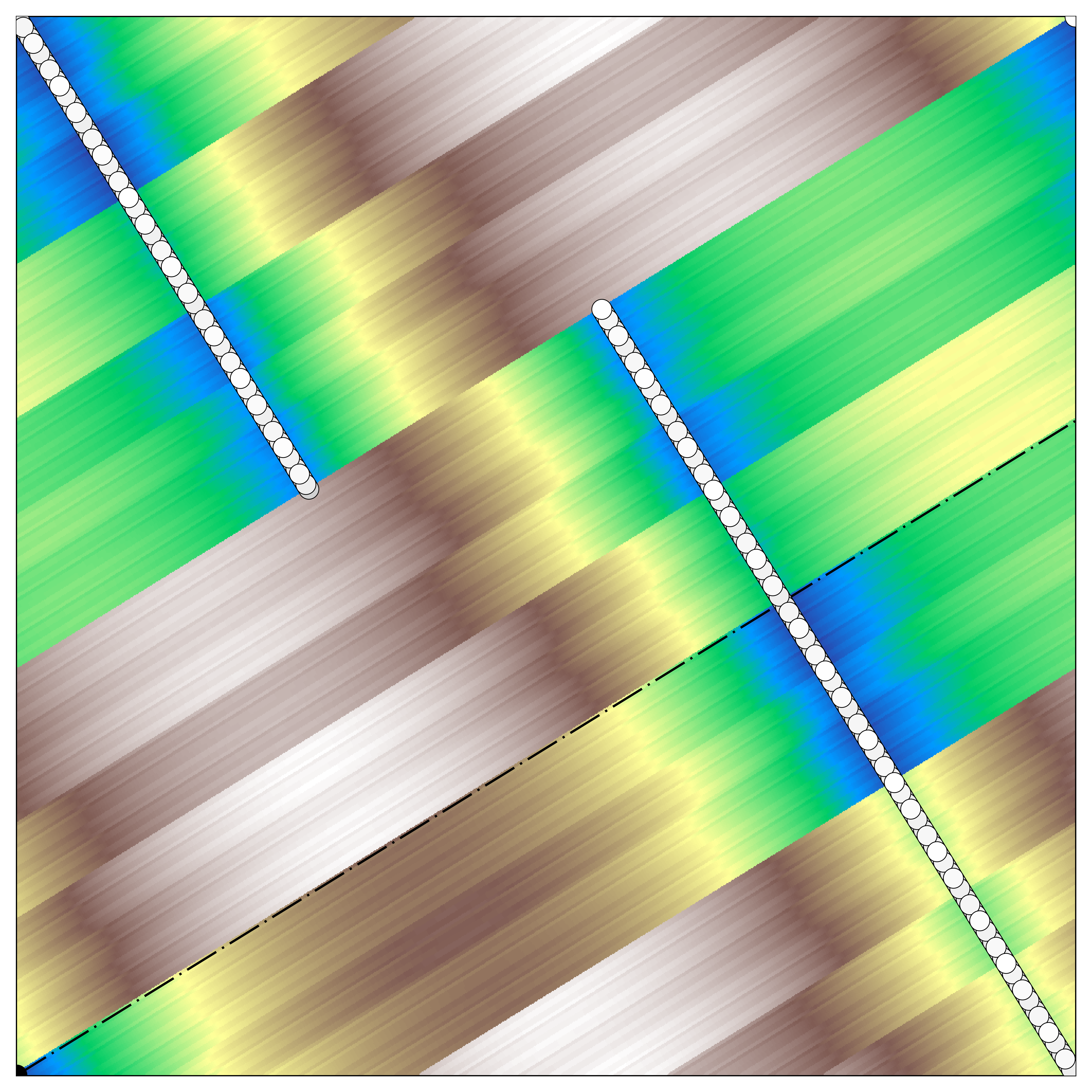}} \\
            \small (b) &
            \raisebox{-0.95\height}{\includegraphics[width=0.35\textwidth]{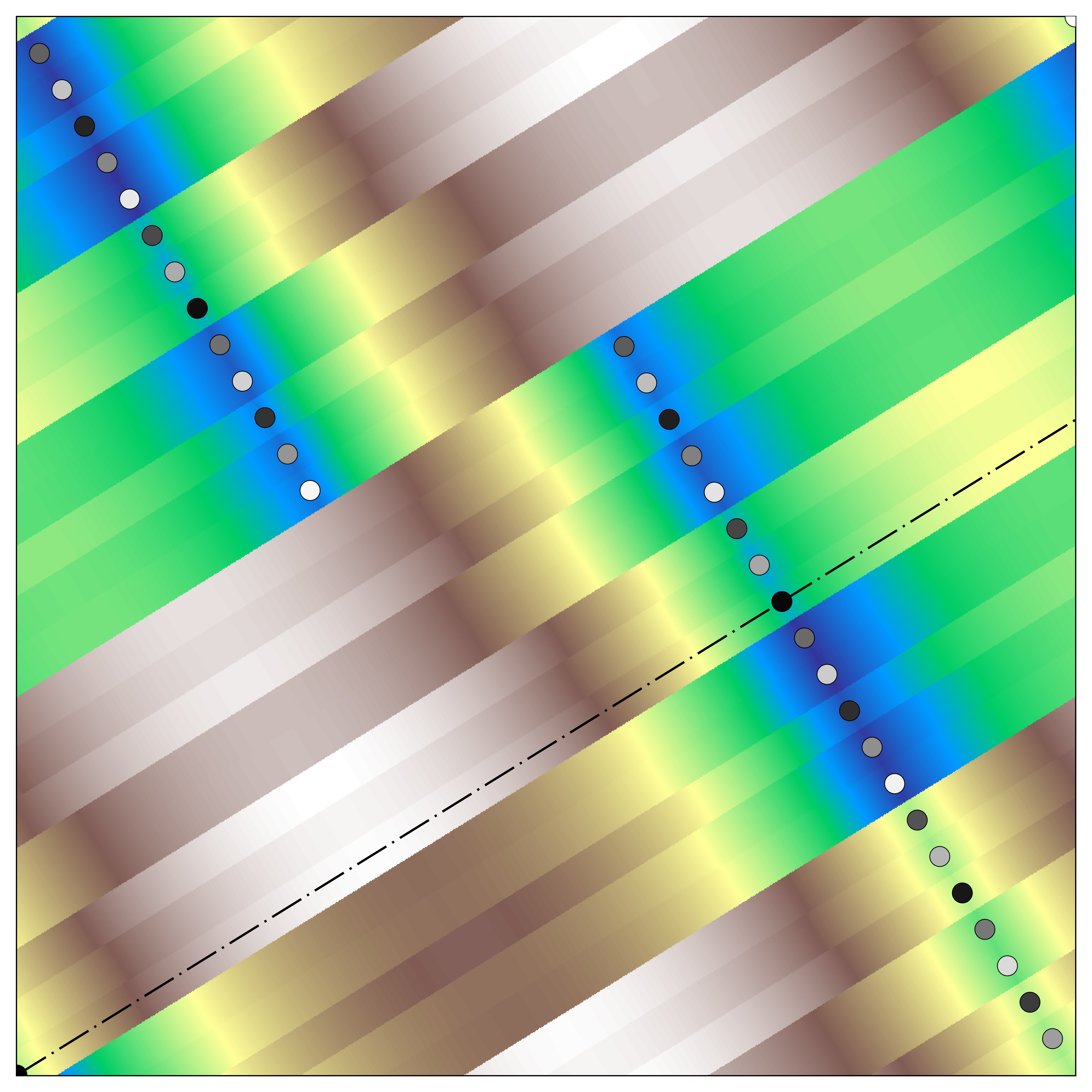}} \\
            \small (c) &
            \raisebox{-0.95\height}{\includegraphics[width=0.35\textwidth]{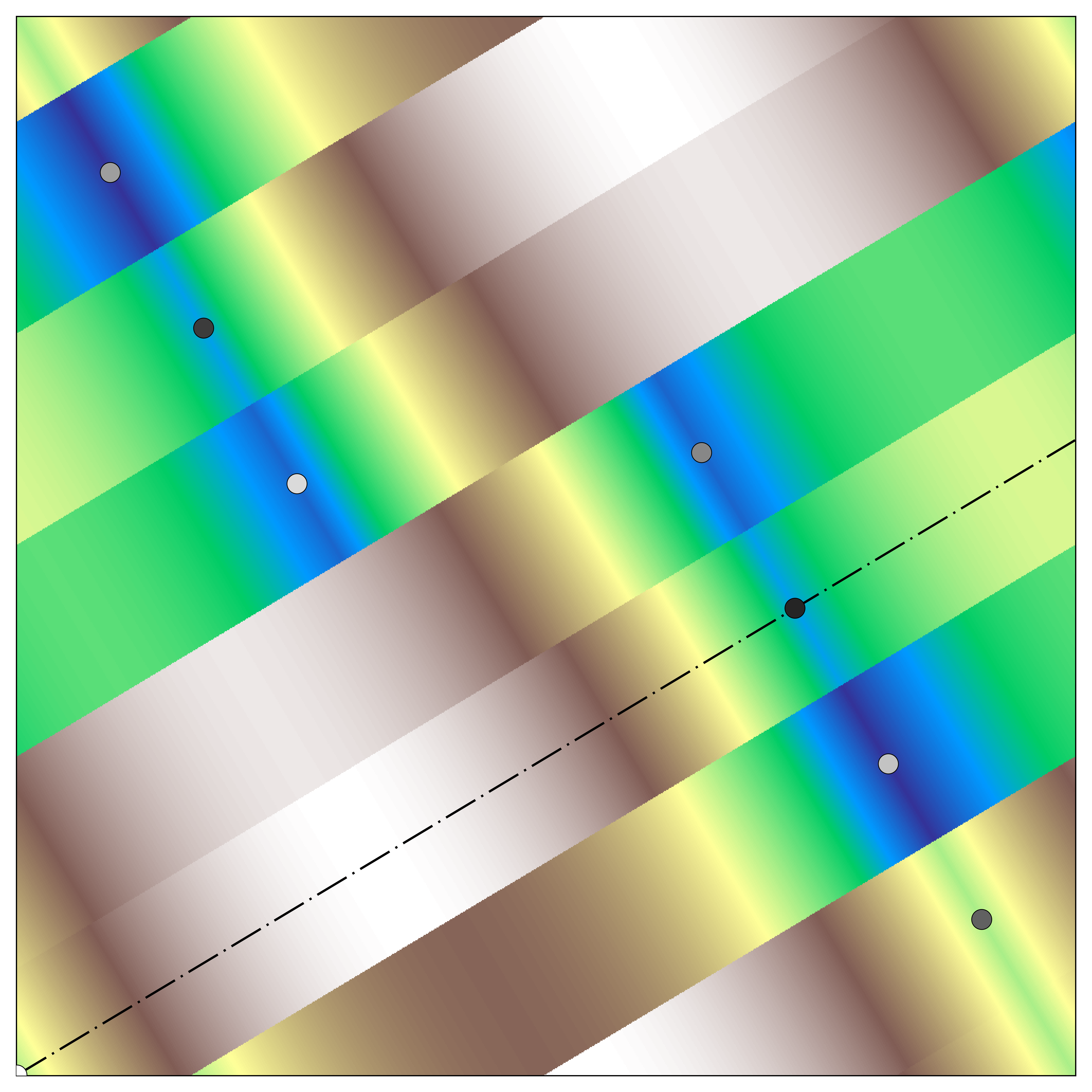}}
        \end{tabular}
    \end{center}
    \caption{ Potentials for the HS of the Fibonacci QC and its approximants. (a) The potential corresponding to the full Fibonacci QC. The abrupt discontinuities at each end of the atomic arcs radiate out as finer and finer discontinuities along the dashed line $\mathbb L$ in either direction. (b) The potential corresponding to the approximant with slope $13/21$ possessing $13+21=34$ atoms in the crystalline unit cell. (c) The potential corresponding to the approximant with slope $3/5$ containing $3+5=8$ atoms in the crystalline unit cell.}
    \label{fig:potential}
\end{figure}

Second, we show that the C+P approach is strictly more powerful than the method of approximants, by analyzing the approximants with the methods of our C+P approach (compare Fig.~\ref{fig:potential}). We resolve  issues that arise in connection with the specification of the potential term, and that concern convergence to the solution of the promoted QC differential equations.

Third, we give a new concrete formulation of DFT for quasicrystals in the cut-and-project method by promoting the electron density to a free parameter in the DFT Lagrangian. This facilitates the direct calculation of QC density of states (DOS) without resort to the approximant method for many-electron DFT. In particular, the DFT++ formulation \cite{DFTpp} of DFT is seen to avoid the difficulties with non-local particle interactions that were encountered when previous C+P approaches were applied directly to DFT \cite{paper-preq-important}. Our C+P method applies to electronic, phononic, magnonic, and photonic systems, and indeed extends to general cross-exchange-correlation functional theory, making this approach universal for any quasi-symmetric physical system. We illustrate our methods on the electronics of a Fibonacci QC, before expanding their applicability to more general quasicrystals and physical theories.

\begin{figure*}
\includegraphics[scale=0.55]{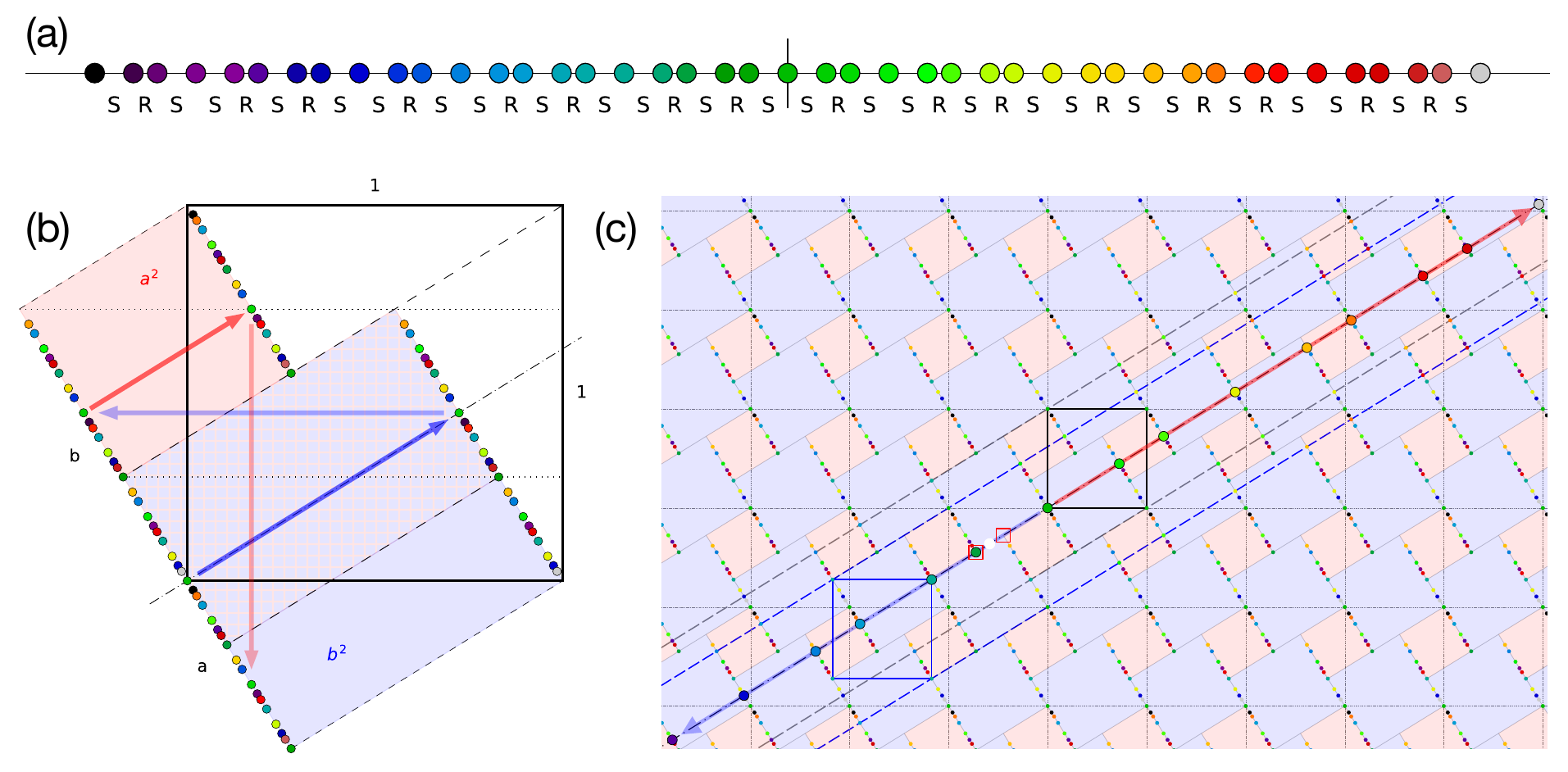}
\caption{The Fibonacci QC in both physical space (PS) and higher space (HS). (a) Construction of the Fibonacci QC in the PS with atoms spaced by either $R$ or $S$, generated by a string substitution rule. The starting seed string is $S$, and the substitutions are $S\rightarrow SR$ and $R\rightarrow S$. 
In this example $S$ and $R$ may denote arbitrary spatial or typological information about the material. 
In the lower figures, (b) illustrates the toric unit cell (as a unit square, or as the union of $a^2$ and $b^2$), while (c) illustrates the translationally invariant HS associated with this description. The red squares have side $a=1/\sqrt{1+\tau^2}$ for the $R$ step, while the blue squares have side $b = \tau / \sqrt{1+\tau^2}$ for the $S$ step. In (b), the cut line $\mathbb L$ of slope $1/\tau=a/b$ wraps infinitely often around the torus without meeting itself. If atoms are placed at distances $a$ and $b$ on $\mathbb L$ according to (a), they will not fill up the square densely, but do densely fill the atomic torus arcs pictured as the two line segments in the square. By taking the closure of these atomic positions, the C+P construction is completed \cite{atoms-as-lines, atoms-as-lines-new}. The unit cell is tiled out to the whole HS plane in (c). Here, the extended line $\mathbb L$ now copies the PS in the plane. The color-coded atoms appear in (a), (b), and (c).}
\label{tab:cut-and-project}
\end{figure*}

We explicitly outline the construction of the Fibonacci quasicrystal  in Fig.~\ref{tab:cut-and-project}. The Fibonacci QC  
consists of a series of atoms with two possible next-neighbor spacings defined in terms of the golden ratio $\tau\simeq1.618$ as either $b = \tau / \sqrt{\tau^2 + 1}\simeq.851$ or $a = 1/\sqrt{\tau^2 + 1}\simeq.526$ \cite{quasicrystal-symm-multifractality}. 
In its classic C+P construction, atoms are placed at the points (with integer coordinates) of the lattice $\mathbb Z^2$ in the 2D plane $\mathbb R^2$, which is the higher space (HS) here. 

The point vectors in the HS are $2\times 1$ column matrices, denoted generically by $\z$. The particular $2\times 1$ matrix $L = [a\ b]^T$ forms the unit vector $\mathbf L$ in the higher space, since $a^2+b^2=1$. Its real scalar multiples $\z = L\x$ form a line $\mathbb L$ in the HS for $\x$ in the PS $\mathbb{R}$. The line $\mathbb L$ has slope $1/\tau$. It represents a faithful copy of the PS, produced by the transformation $L\colon \x\to\z=L\x$.
In the quasicrystal constructed in this copy of the PS, an atom appears projected from the HS lattice $\mathbb Z^2$ to the line under the following conditions:
{(i)} it is above the line, within distance $a$ from it, or 
{(ii)} it is below the line, within distance $b$ from it,
as shown in Fig.~\ref{tab:cut-and-project}(c). 
Thus,  exact coordinates $\x$ for the atomic locations in the PS are determined.

In fact, an equivalent construction approach \cite{atoms-as-lines} is adopted here, in order to facilitate the subsequent determination of the appropriate atomic potentials in the HS.  Rather than projecting the atoms from $\mathbb Z^2$ to $\mathbb L$, elongate each atom point in $\mathbb Z^2$ to its \emph{atomic segment}, a line segment perpendicular to $\mathbb L$ and containing the atom point, but stretching down for a distance of $a$ and up for a distance of $b$.  This produces an infinite number of segments in the HS, each of length $a+b$. 
An atom then appears in the quasicrystal if its atomic segment intersects $\mathbb L$.

Next, we present the construction of the Fibonacci QC potential in the HS. In the Born-Oppenheimer description of a material in its natural space, the atoms are taken as fixed points, each generating a potential with which electrons interact. 

For the Fibonacci QC in the PS, the total potential $V(\x)$ along the line $\mathbb L$ is the sum $V(\x) = \sum_{\x_a} \widetilde V(\x - \x_a)$ of the individual atomic potential $\widetilde V(\x)$ at each atomic position $\x_a$. 
This sum may be written as 
$V(\x) = \delta_{\{\x_a\}} * \tilde{V}(\x)$, 
convolving the generic atom potential with the indicator function of the set $\{\x_a\}$ of atom point locations in the Fibonacci QC. 

For the Fibonacci QC in the HS, the Born-Oppenheimer description of the quasicrystal requires modification. In the higher-dimensional physics, the quasicrystal atoms  are represented by their entire atomic segments. The atomic potential is given by
$$
\widetilde V(\z)=
\begin{cases}
\widetilde V(\x) &\mbox{if } \z\in\mathbb L\,;\\
0 & \mbox{otherwise}.
\end{cases}
$$
Then, writing $\{\z_a\}$ for the set of points in the atomic segment, let $V(\z) = \delta_{\{\z_a\}} * \widetilde V (\z)$,
convolving the generic HS atom potential $\widetilde V (\z)$ with the indicator function of the set $\{\z_a\}$ of atomic segment locations in the Fibonacci QC.

The potential
$V(\z)$ 
respects the translational symmetry of the HS, and restricting $V(\z)$ to the cut line given by $\mathbb L$ reproduces $V(\x)$. The potential $V(\z)$ for the Fibonacci QC is pictured in Fig.~\ref{fig:potential}(a). Note that this potential is continuous along $\mathbb L$ and discontinuous along $\mathbb L^\perp$.

Generalizations of simple theories of physics have been considered in their extension to the HS. Take the Schr\"odinger equation for the QC \cite{most-important-paper}. We can write the QC formulation of the Schr\"odinger equation compactly. Thus, in HS the form from \cite{most-important-paper} becomes
\begin{align*}
\left(-\frac{\hbar^2}{2m}(\mathbf L\cdot \nabla)^2 + V(\z)\right)\psi(\z) = E \psi(\z).
\end{align*}
Since the HS is a torus, the physical intuition here is that the 1D PS line is wrapped into the square toroidal unit cell in the HS. As all operators are geometrically constrained to only include information along the implentation of the PS into the HS as $\mathbb L$, the differential equation is identical.

Suppose we wish to use DFT with the potential $V(\z)$. A common issue in extending DFT to the HS comes from the existence of the nonlinear term $e\int \frac{n(\x')}{|\x-\x'|}\,d\x'$, used for computing electron-electron interaction. Computational rather than physical, this equation is non-local in nature and will not give rise to a natural equation in the HS. In order to generalize DFT, a local version is required. This can be made by promoting the potential to a first order variable, as is done in \cite{DFTpp} to effect fast DFT algorithmic convergence as: $\mathcal{L}_{\text{LDA}} \left( \{\psi_i(\x)\}, \phi(\x) \right) =$
\begin{align*}
& -\frac{\hbar^2}{2m_e} \sum_i f_i \int d\x \, \psi_i^*(\x) \nabla^2 \psi_i(\x) \\
&\quad + \int d\x \, V(\x) n(\x) + \int d\x \, \epsilon_{\text{xc}}(n(\x)) n(\x) \\
&\quad - \int d\x \, \phi(\x) (n(\x) - n_0) - \frac{\epsilon_0}{2} \int d\x \, ||\nabla \phi(\x)||^2.
\end{align*}

The form of DFT has been extensively described elsewhere \cite{martin2020electronic}. Here $\phi$ is the electron potential, on the same order as the electron wavefunctions $\psi_i$ and evolved with them. This encodes the non-local term. $\epsilon_{\text{xc}}$ is an unspecified exchange correlation and $n$ is electron density. This equation evidently applies to the PS. Extending the electron density $n(\x)$ from PS to HS is unproblematic, as it is defined pointwise.

As with the generalization of the Schr\"odinger equation, all terms may be geometrically restricted to operate with information along $\mathbb L$ in the HS, and so have the same solution as in the PS: $\mathcal{L}_{\text{LDA}} \left( \{\psi_i(\z)\}, \phi(\z) \right) = $
\begin{align*}
& -\frac{\hbar^2}{2m_e} \sum_i f_i \int d^2\z \, \psi_i^*(\z) (\mathbf L\cdot \nabla)^2 \psi_i(\z) \\
&\quad + \int d^2\z \, V(\z) n(\z)  + \int d^2\z \, \epsilon_{\text{xc}}(n(\z)) n(\z) \\
&\quad - \int d^2\z \, \phi(\z) (n(\z) - n_0) - \frac{\epsilon_0}{2} \int d^2\z \, ||\mathbf L\cdot\nabla \phi(\z)||^2.
\end{align*}

Of special note are the integrals, which, rather than operating over $\mathbb{R}$ now operate over $\mathbb{R}^2$. %
It is important to note that this new equation may be taken as a purely geometric restatement of the PS DFT over the line $\mathbb L$. Supposing that the integrals were restricted to $\mathbb L$ via a Dirac delta, this would be a trivial restatement. The additional physical intuition that is enabled is first realized by allowing the integrals to exist over the entire HS, and then noting that, since the HS has translational symmetry, this HS DFT appears as a crystalline atomic DFT over the HS primitive cell. In order to complete this picture, before continuing to theoretical issues, the density of state (DOS) is presented in terms of the HS to indicate concretely how physical information is directly encoded in the HS.

The unit cell for quasi-crystals is not well defined in the PS; however, the DOS is still meaningful, so a simple computational method for the DOS is determined. Because the DOS has no spatial information, it trivially respects the spatial symmetry of a quasi-crystal.
The standard expression for the density of states,
$$
\rho(\omega) = \sum_{\mu} \int \left(\frac{dk}{2\pi}\right)^d \, \delta\big(\omega - \epsilon_\mu(k)\big),
$$
may be rewritten in the following local form:
$$
\rho(\omega) = \sum_{\mu} \int_{\epsilon_\mu(k) = \omega} \frac{dS}{|\nabla_k \epsilon_\mu(k)|}.
$$

Now, for QCs, in the HS, the primitive cell does exist, and thus the reciprocal space to the HS is identical to that for standard crystals. The geometric relation between the HS and the PS

The mathematical apparatus in the subsequent sections gives a precise way to determine the relation between the DOS in the HS and the DOS in the PS. This is given by:
$$
\rho(\omega) = \sum_{\mu} \int_{\epsilon_\mu(k) = \omega} \frac{dS}{|(\mathbf L\cdot \nabla_{k}) \epsilon_\mu(k)|}
$$

Here, the integration takes place over the QC primitive cell. Given the C+P construction of the Fibonacci QC, corresponding atomic potentials have been derived in HS, as well as generalizations of DFT and the DOS. An exact calculation of the DOS of the Fibonacci QC is enabled, without resorting to approximants through the use of this theory.

Now we address several of the underlying mathematical procedures. First is the general procedure connecting the PS and HS, which enables local physical theories to be modified to the HS. Then, approximants are dealt with, and used to demonstrate that converting PS integrals to HS integrals is well defined. Finally, some practical statements about the nature of the theory are made to make it computationally viable.

For simplicity, the materials treated are assumed to be of a single atomic element. This enables simple formulae, which can be generalized to multi-atomic theories with relative ease. A QC material with a C+P construction is composed of a set of atoms $A$ in PS, which are related to a set of atoms $\uparrow A$ in HS by a transform $L$. This $L$ is an $n\times d$ matrix composed of $d$ orthonormal vectors in $\mathbb{R}^n$. Additionally, the HS is equipped with a complete set of vectors defining translation symmetry $\{t_i\}_{i=1}^n$. The position of an atom in PS is $\x_a$ for $a \in A$ and the position of an atom in HS is $\z_a$ for $a \in \uparrow A$ which must respect the translational symmetry of the HS. $A$ and $\uparrow A$ have the property that $\z_a = L  \x_a$, which embeds the atoms in the PS to a coordinate hyperplane in HS. Thus $A$ is a subset of $\uparrow A$. In the case that the material may be constructed by classical C+P, the HS will contain atomic arcs perpendicular to the image hyperplane of $L$. However, the C+P construction just described is broadened, and may be used with many configurations in HS to model perturbations or nonstandard QCs \cite{atoms-as-lines}. Simple requirements on the PS to ensure the material is realistic include having a countable infinity of atoms which nowhere get arbitrarily close to each other (i.e., have no accumulation points).

In order to consider the PS and HS spaces as dual in a formal sense, $\uparrow$ and $\downarrow$ are introduced. PS is considered the physical space, parametrized by $\x$. $\lps$ is the embedding of PS in the HS, shorthanded as $\mathbb L$, and parametrized by $\z$, restricted to $\z = L\x$. HS is the full higher space, parametrized by $\z$ without restriction. The properties of the C+P construction center on the translation symmetry of the $\hs$. Let $\Uz$ be $\z$ modulo the $t_i$ so it may be expressed as a linear combination $\sum_{i=1}^n a_i \bm{t}_i$ for $a_i \in [0,1)$ so that $\Uz = \U{\z + \bm{t}_i}$. While HS is $\mathbb{R}^n$, the image of the $\U{\phantom{n}}$ operation gives $\Uhs$ as an $n$-dimensional torus, forming the primitive cell. For a material to be a QC, the $\Ulps$ of the C+P construction must be dense in the unit cell. 
While here, density is simply assumed, it may be specified by geometric constraints on $L$ and the $t_i$ with the use of the Kronecker density theorem \cite{kronecker}. For any space $S$, the  space of functions from $S$ to $\mathbb{C}$ is denoted $S^\dagger$. Then, $\uparrow\colon\ps\to\C{\Ulps}$ may be defined as the spatial lifting from the $\ps$ to $\hs$, and the dual $\downarrow\colon\Uhs^\dagger\to\ps^\dagger$ converts a function on $\Uhs$ to a function on $\ps$. Their definitions are summarized by the following diagram which serves as a general template for how physical meanings are translated between the $\ps$ and $\hs$:
\begin{equation}
\begin{split}    
\begin{aligned}
&\ps~~ \overbrace{
  \xrightarrow{L} ~\lps \xrightarrow{\mathit{mod\,by\,t_i}} \Ulps \xrightarrow{\mathit{closure}}
}^{(\mathrm{lifting})\uparrow} ~\C{\Ulps} & \\
&\uparrow\flat~~~~~~~~~~~~~~~~~~~~~~~~~~~~~~~~~~~~~~~~~~~~~~\downarrow\# &. \\
&\ps^\dagger~\underbrace{\xleftarrow[L^{\dagger}]{}\lps^\dagger \xleftarrow[\mathit{restrict}]{}\hs^\dagger\xleftarrow[\mathit{tile\,by\,t_i}]{}}_{(\mathrm{lowering})\downarrow}~\Uhs^\dagger &
\end{aligned}
\end{split}    
\end{equation}

Here the $\# : S \rightarrow S^\dagger$ and $\flat : S^\dagger \rightarrow S$ refer respectively to creating the function space over a geometrical space $S$, and to recovering the geometrical space as the ideal space $\mathrm{Spec}\,S^\dagger$ of its functional space $S^\dagger$. The $\uparrow$ refers to an operation on a subset of $\ps$, first implanting it with $L$, then moving the subset to the primitive cell, taking the modulus with respect to $T_i$, and finally taking the closure. Note that this can be extended to the indicator function as $\uparrow 1_{A} = 1_{\uparrow A}$ for $A \subseteq \ps$ taking the standard definition that $1_{A}$ is $1$ for points in $A$ and $0$ otherwise. The $\downarrow$ is defined for a function $f : \Uhs \rightarrow \mathbb{Z}$. First, we take the periodic continuation of $f$ over $\hs$ as $\U{f}(z) = f(\Uz)$ so it is defined on $\hs$. Then we restrict it to $\lps$ so that it's definition on $\ps$ may be found as $f(x) = \U{f}(Lx)$.  Note that $\uparrow \downarrow 1_{A} = 1_{A}$. Now, we find the exact mathematical relation between $A$ and $\uparrow A$ as $\uparrow A = \overline{\{\U{\z_i} : \x_a, a \in \mathrm{A}\}}$. The interpretation of this, is that all the atoms are individually raised to the $\hs$, then the closure is used to complete the atoms from $\Ulps$ to $\hs$. For QCs described by the classical C+P construction, this lift operation will create well-defined forms. While an atomic collection $A$ could be varied independently from the choice of $L$, potentially making the atomic placement in the $\hs$ visually random, for materials defined by their pure point diffraction or other representations as discussed both in Fig.~\ref{tab:cut-and-project} and the supplemental materials, this operation will create natural forms orthogonal to the $\lps$ subspace. Alternatively, one could start with arbitrary forms respecting the $\hs$ translation symmetry such that their intersection with $\lps$ is always a single point. Term this collection $\uparrow A$. Then, computing $\downarrow (\uparrow A)$ is simply taking the intersection of the embedding $\lps$ with $(\uparrow A)$. Note that for any C+P construction, it must be that $\downarrow\uparrow A = A$ for the physical equations discussed to generalize to the HS. 

Note that, as $\Ulps$ is merely dense in $\Uhs$, not all $\z$ have corresponding $\x$, so defining $\downarrow z$ directly cannot work. In general the lifting of a function to the HS is not uniquely defined. If the potential $V(\z)$ in HS were to be continuous and restrict exactly to the atomic potential as $\downarrow V(\z) = V(\x)$, the construction would not necessitate the consideration of atomic pointsets. For instance, given the QC potential $V(\x) = \sin(\frac{\x}{2\pi}) + \sin(\frac{\sqrt{2}\x}{2\pi})$. Let $T_1 = (1, 0)$, $T_2 = (0, \sqrt{2})$ with $L = (1/\sqrt{2}, 1/\sqrt{2})^T$. Then, for $\z = (z_1, z_2)$, $\uparrow V = (\sin(\frac{z_1}{2\pi}),\,\sin(\frac{\sqrt{2}z_2}{2\pi}))$ exists as the direct implantation of $V$ in $\hs$. Note that $\downarrow \uparrow V = V$ as desired.

However, both for the Fibonacci QC, and in general, the HS potential will not be continuous. We say that a potential in HS is compatible with one in PS if $\downarrow V(\z) = V(\x)$. With this in view, the Fibonacci QC HS potential construction generalizes to all C+P constructions and is compatible with the potential in PS.

From the lifting and lowering notation, this method is extremely general. Given a potential $V$ in PS, the previous procedure allows the lift of V, $\uparrow V$, to be computed. Suppose we wish to calculate $\nabla V$. Rather than calculate it directly in PS, it may be calculated as $\downarrow L\cdot \nabla \uparrow V$, where $L\cdot \nabla$ is the lifting of the $\nabla$ operator. With this in mind, local physical theories may be lifted to HS, computations may be done there, then lowered. For instance, with the path integral formulation:
$$\phi(\x, t) = \frac{1}{Z} \int_{\x(0)=\x}\mathcal{D}\x e^{iS[\x, \dot{\x}]},$$
The generalization of this integral to the QC HS system is:
$$\phi(\z, t) = \frac{1}{Z} \int_{\z(0)=\z}\mathcal{D}\x e^{iS[\z, \dot{\z}_{\parallel L}]}.$$
Note that the equivalence here is \textit{geometric}. Observables are defined dually as the PS-quantity derived from the HS-solution.

\subsection{Conclusion}

By extending the concept of the Schr\"odinger equation and DFT to QCs, we have demonstrated how standard electronic structure methods 
extend to systems lacking periodic translational symmetry,
providing a natural way to study electronic states and their evolution within a QC. Importantly, the projection of the HS Schr\"odinger equation onto physical space ensures that the fundamental quantum mechanical properties remain intact. The adaptation of DFT required a reconsideration of the electron-electron interaction term, but a DFT++ formulation preserves the essential features of the Coulomb potential in QC systems, with tractable computational features. By mapping QCs to a higher-dimensional parent crystal and deriving the DOS from the associated band structure, we establish here a method that allows for efficient, direct numerical calculations. This approach ensures that the unique spectral properties of QCs are faithfully captured, while maintaining a direct correspondence with traditional crystal calculations.

\section*{Acknowledgment}

This work was supported as part of the Center for Energy Efficient Magnonics, an Energy Frontier Research Center funded by the U.S.\@ Department of Energy, Office of Science, Basic Energy Sciences, under Award number DE-AC02-76SF00515. T.K.\@ acknowledges the support provided from the Ames National Laboratory, the U.S.\@ Department of Energy, Office of Science, Basic Energy Sciences, Materials Science and Engineering Division, under contract No.\@ DE-AC02-07CH11358.

\bibliography{citations}%

\begin{thebibliography}{28}%
\makeatletter
\providecommand \@ifxundefined [1]{%
 \@ifx{#1\undefined}
}%
\providecommand \@ifnum [1]{%
 \ifnum #1\expandafter \@firstoftwo
 \else \expandafter \@secondoftwo
 \fi
}%
\providecommand \@ifx [1]{%
 \ifx #1\expandafter \@firstoftwo
 \else \expandafter \@secondoftwo
 \fi
}%
\providecommand \natexlab [1]{#1}%
\providecommand \enquote  [1]{``#1''}%
\providecommand \bibnamefont  [1]{#1}%
\providecommand \bibfnamefont [1]{#1}%
\providecommand \citenamefont [1]{#1}%
\providecommand \href@noop [0]{\@secondoftwo}%
\providecommand \href [0]{\begingroup \@sanitize@url \@href}%
\providecommand \@href[1]{\@@startlink{#1}\@@href}%
\providecommand \@@href[1]{\endgroup#1\@@endlink}%
\providecommand \@sanitize@url [0]{\catcode `\\12\catcode `\$12\catcode `\&12\catcode `\#12\catcode `\^12\catcode `\_12\catcode `\%12\relax}%
\providecommand \@@startlink[1]{}%
\providecommand \@@endlink[0]{}%
\providecommand \url  [0]{\begingroup\@sanitize@url \@url }%
\providecommand \@url [1]{\endgroup\@href {#1}{\urlprefix }}%
\providecommand \urlprefix  [0]{URL }%
\providecommand \Eprint [0]{\href }%
\providecommand \doibase [0]{https://doi.org/}%
\providecommand \selectlanguage [0]{\@gobble}%
\providecommand \bibinfo  [0]{\@secondoftwo}%
\providecommand \bibfield  [0]{\@secondoftwo}%
\providecommand \translation [1]{[#1]}%
\providecommand \BibitemOpen [0]{}%
\providecommand \bibitemStop [0]{}%
\providecommand \bibitemNoStop [0]{.\EOS\space}%
\providecommand \EOS [0]{\spacefactor3000\relax}%
\providecommand \BibitemShut  [1]{\csname bibitem#1\endcsname}%
\let\auto@bib@innerbib\@empty
%</preamble>
\bibitem [{\citenamefont {Shechtman}\ \emph {et~al.}(1984)\citenamefont {Shechtman}, \citenamefont {Blech}, \citenamefont {Gratias},\ and\ \citenamefont {Cahn}}]{qc-discovery}%
  \BibitemOpen
  \bibfield  {author} {\bibinfo {author} {\bibfnamefont {D.}~\bibnamefont {Shechtman}}, \bibinfo {author} {\bibfnamefont {I.}~\bibnamefont {Blech}}, \bibinfo {author} {\bibfnamefont {D.}~\bibnamefont {Gratias}},\ and\ \bibinfo {author} {\bibfnamefont {J.~W.}\ \bibnamefont {Cahn}},\ }\bibfield  {title} {\bibinfo {title} {Metallic phase with long-range orientational order and no translational symmetry},\ }\href@noop {} {\bibfield  {journal} {\bibinfo  {journal} {Phys. Rev. Lett.}\ }\textbf {\bibinfo {volume} {53}},\ \bibinfo {pages} {1951} (\bibinfo {year} {1984})}\BibitemShut {NoStop}%
\bibitem [{\citenamefont {Dubost}\ \emph {et~al.}(1986)\citenamefont {Dubost}, \citenamefont {Lang}, \citenamefont {Tanaka}, \citenamefont {Sainfort},\ and\ \citenamefont {Audier}}]{history-2}%
  \BibitemOpen
  \bibfield  {author} {\bibinfo {author} {\bibfnamefont {B.}~\bibnamefont {Dubost}}, \bibinfo {author} {\bibfnamefont {J.}~\bibnamefont {Lang}}, \bibinfo {author} {\bibfnamefont {M.}~\bibnamefont {Tanaka}}, \bibinfo {author} {\bibfnamefont {P.}~\bibnamefont {Sainfort}},\ and\ \bibinfo {author} {\bibfnamefont {M.}~\bibnamefont {Audier}},\ }\bibfield  {title} {\bibinfo {title} {Large alculi single quasicrystals with triacontahedral solidification morphology},\ }\href@noop {} {\bibfield  {journal} {\bibinfo  {journal} {Nature}\ }\textbf {\bibinfo {volume} {324}},\ \bibinfo {pages} {48} (\bibinfo {year} {1986})}\BibitemShut {NoStop}%
\bibitem [{\citenamefont {Kraus}\ \emph {et~al.}(2012)\citenamefont {Kraus}, \citenamefont {Lahini}, \citenamefont {Ringel}, \citenamefont {Verbin},\ and\ \citenamefont {Zilberberg}}]{QC-are-cool-1}%
  \BibitemOpen
  \bibfield  {author} {\bibinfo {author} {\bibfnamefont {Y.~E.}\ \bibnamefont {Kraus}}, \bibinfo {author} {\bibfnamefont {Y.}~\bibnamefont {Lahini}}, \bibinfo {author} {\bibfnamefont {Z.}~\bibnamefont {Ringel}}, \bibinfo {author} {\bibfnamefont {M.}~\bibnamefont {Verbin}},\ and\ \bibinfo {author} {\bibfnamefont {O.}~\bibnamefont {Zilberberg}},\ }\bibfield  {title} {\bibinfo {title} {Topological states and adiabatic pumping in quasicrystals},\ }\href@noop {} {\bibfield  {journal} {\bibinfo  {journal} {Phys. Rev. Lett.}\ }\textbf {\bibinfo {volume} {109}},\ \bibinfo {pages} {106402} (\bibinfo {year} {2012})}\BibitemShut {NoStop}%
\bibitem [{\citenamefont {Kraus}\ \emph {et~al.}(2013)\citenamefont {Kraus}, \citenamefont {Ringel},\ and\ \citenamefont {Zilberberg}}]{paper-inspiration}%
  \BibitemOpen
  \bibfield  {author} {\bibinfo {author} {\bibfnamefont {Y.~E.}\ \bibnamefont {Kraus}}, \bibinfo {author} {\bibfnamefont {Z.}~\bibnamefont {Ringel}},\ and\ \bibinfo {author} {\bibfnamefont {O.}~\bibnamefont {Zilberberg}},\ }\bibfield  {title} {\bibinfo {title} {Four-dimensional quantum hall effect in a two-dimensional quasicrystal},\ }\href@noop {} {\bibfield  {journal} {\bibinfo  {journal} {Phys. Rev. Lett.}\ }\textbf {\bibinfo {volume} {111}},\ \bibinfo {pages} {226401} (\bibinfo {year} {2013})}\BibitemShut {NoStop}%
\bibitem [{\citenamefont {Bandres}\ \emph {et~al.}(2016)\citenamefont {Bandres}, \citenamefont {Rechtsman},\ and\ \citenamefont {Segev}}]{topological-photonic-quasicrystals}%
  \BibitemOpen
  \bibfield  {author} {\bibinfo {author} {\bibfnamefont {M.~A.}\ \bibnamefont {Bandres}}, \bibinfo {author} {\bibfnamefont {M.~C.}\ \bibnamefont {Rechtsman}},\ and\ \bibinfo {author} {\bibfnamefont {M.}~\bibnamefont {Segev}},\ }\bibfield  {title} {\bibinfo {title} {Topological photonic quasicrystals: fractal topological spectrum and protected transport},\ }\href@noop {} {\bibfield  {journal} {\bibinfo  {journal} {Phys. Rev. X}\ }\textbf {\bibinfo {volume} {6}},\ \bibinfo {pages} {011016} (\bibinfo {year} {2016})}\BibitemShut {NoStop}%
\bibitem [{\citenamefont {Tamura}\ \emph {et~al.}(2025)\citenamefont {Tamura}, \citenamefont {Abe}, \citenamefont {Yoshida}, \citenamefont {Shimozaki}, \citenamefont {Suzuki}, \citenamefont {Ishikawa}, \citenamefont {Labib}, \citenamefont {Avdeev}, \citenamefont {Kinjo}, \citenamefont {Nawa} \emph {et~al.}}]{antiferromagnetic-QC}%
  \BibitemOpen
  \bibfield  {author} {\bibinfo {author} {\bibfnamefont {R.}~\bibnamefont {Tamura}}, \bibinfo {author} {\bibfnamefont {T.}~\bibnamefont {Abe}}, \bibinfo {author} {\bibfnamefont {S.}~\bibnamefont {Yoshida}}, \bibinfo {author} {\bibfnamefont {Y.}~\bibnamefont {Shimozaki}}, \bibinfo {author} {\bibfnamefont {S.}~\bibnamefont {Suzuki}}, \bibinfo {author} {\bibfnamefont {A.}~\bibnamefont {Ishikawa}}, \bibinfo {author} {\bibfnamefont {F.}~\bibnamefont {Labib}}, \bibinfo {author} {\bibfnamefont {M.}~\bibnamefont {Avdeev}}, \bibinfo {author} {\bibfnamefont {K.}~\bibnamefont {Kinjo}}, \bibinfo {author} {\bibfnamefont {K.}~\bibnamefont {Nawa}}, \emph {et~al.},\ }\bibfield  {title} {\bibinfo {title} {Observation of antiferromagnetic order in a quasicrystal},\ }\href@noop {} {\bibfield  {journal} {\bibinfo  {journal} {Nat. Phys.}\ }\textbf {\bibinfo {volume} {21}},\ \bibinfo {pages} {974} (\bibinfo {year} {2025})}\BibitemShut {NoStop}%
\bibitem [{\citenamefont {Hao}\ \emph {et~al.}(2024)\citenamefont {Hao}, \citenamefont {Zhan}, \citenamefont {Pantale{\'o}n}, \citenamefont {He}, \citenamefont {Zhao}, \citenamefont {Watanabe}, \citenamefont {Taniguchi}, \citenamefont {Guinea},\ and\ \citenamefont {He}}]{advanced-moire-experiment-bands}%
  \BibitemOpen
  \bibfield  {author} {\bibinfo {author} {\bibfnamefont {C.-Y.}\ \bibnamefont {Hao}}, \bibinfo {author} {\bibfnamefont {Z.}~\bibnamefont {Zhan}}, \bibinfo {author} {\bibfnamefont {P.~A.}\ \bibnamefont {Pantale{\'o}n}}, \bibinfo {author} {\bibfnamefont {J.-Q.}\ \bibnamefont {He}}, \bibinfo {author} {\bibfnamefont {Y.-X.}\ \bibnamefont {Zhao}}, \bibinfo {author} {\bibfnamefont {K.}~\bibnamefont {Watanabe}}, \bibinfo {author} {\bibfnamefont {T.}~\bibnamefont {Taniguchi}}, \bibinfo {author} {\bibfnamefont {F.}~\bibnamefont {Guinea}},\ and\ \bibinfo {author} {\bibfnamefont {L.}~\bibnamefont {He}},\ }\bibfield  {title} {\bibinfo {title} {Robust flat bands in twisted trilayer graphene moir{\'e} quasicrystals},\ }\href@noop {} {\bibfield  {journal} {\bibinfo  {journal} {Nat. Commun.}\ }\textbf {\bibinfo {volume} {15}},\ \bibinfo {pages} {8437} (\bibinfo {year} {2024})}\BibitemShut {NoStop}%
\bibitem [{\citenamefont {Baake}\ and\ \citenamefont {Grimm}(2020)}]{thue-morse}%
  \BibitemOpen
  \bibfield  {author} {\bibinfo {author} {\bibfnamefont {M.}~\bibnamefont {Baake}}\ and\ \bibinfo {author} {\bibfnamefont {U.}~\bibnamefont {Grimm}},\ }\bibfield  {title} {\bibinfo {title} {Inflation versus projection sets in aperiodic systems: the role of the window in averaging and diffraction},\ }\href@noop {} {\bibfield  {journal} {\bibinfo  {journal} {Acta Crystallogr. A}\ }\textbf {\bibinfo {volume} {76}},\ \bibinfo {pages} {559} (\bibinfo {year} {2020})}\BibitemShut {NoStop}%
\bibitem [{\citenamefont {Janssen}(1986)}]{QCsymmetry}%
  \BibitemOpen
  \bibfield  {author} {\bibinfo {author} {\bibfnamefont {T.}~\bibnamefont {Janssen}},\ }\bibfield  {title} {\bibinfo {title} {Crystallography of quasi-crystals},\ }\href@noop {} {\bibfield  {journal} {\bibinfo  {journal} {Acta Crystallogr. A}\ }\textbf {\bibinfo {volume} {42}},\ \bibinfo {pages} {261} (\bibinfo {year} {1986})}\BibitemShut {NoStop}%
\bibitem [{\citenamefont {Yamamoto}(1996)}]{QCrystallographyGuide}%
  \BibitemOpen
  \bibfield  {author} {\bibinfo {author} {\bibfnamefont {A.}~\bibnamefont {Yamamoto}},\ }\bibfield  {title} {\bibinfo {title} {Crystallography of quasiperiodic crystals},\ }\href@noop {} {\bibfield  {journal} {\bibinfo  {journal} {Acta Crystallogr. A}\ }\textbf {\bibinfo {volume} {52}},\ \bibinfo {pages} {509} (\bibinfo {year} {1996})}\BibitemShut {NoStop}%
\bibitem [{\citenamefont {Takakura}\ \emph {et~al.}(2007)\citenamefont {Takakura}, \citenamefont {Gomez}, \citenamefont {Yamamoto}, \citenamefont {De~Boissieu},\ and\ \citenamefont {Tsai}}]{firstAtomicStructure}%
  \BibitemOpen
  \bibfield  {author} {\bibinfo {author} {\bibfnamefont {H.}~\bibnamefont {Takakura}}, \bibinfo {author} {\bibfnamefont {C.~P.}\ \bibnamefont {Gomez}}, \bibinfo {author} {\bibfnamefont {A.}~\bibnamefont {Yamamoto}}, \bibinfo {author} {\bibfnamefont {M.}~\bibnamefont {De~Boissieu}},\ and\ \bibinfo {author} {\bibfnamefont {A.~P.}\ \bibnamefont {Tsai}},\ }\bibfield  {title} {\bibinfo {title} {Atomic structure of the binary icosahedral yb--cd quasicrystal},\ }\href@noop {} {\bibfield  {journal} {\bibinfo  {journal} {Nat. Mater.}\ }\textbf {\bibinfo {volume} {6}},\ \bibinfo {pages} {58} (\bibinfo {year} {2007})}\BibitemShut {NoStop}%
\bibitem [{\citenamefont {Hof}(1995)}]{aperiodic-diffraction}%
  \BibitemOpen
  \bibfield  {author} {\bibinfo {author} {\bibfnamefont {A.}~\bibnamefont {Hof}},\ }\bibfield  {title} {\bibinfo {title} {On diffraction by aperiodic structures},\ }\href@noop {} {\bibfield  {journal} {\bibinfo  {journal} {Commun. Math. Phys.}\ }\textbf {\bibinfo {volume} {169}},\ \bibinfo {pages} {25} (\bibinfo {year} {1995})}\BibitemShut {NoStop}%
\bibitem [{\citenamefont {Arag{\'o}n}\ \emph {et~al.}(2019)\citenamefont {Arag{\'o}n}, \citenamefont {Naumis},\ and\ \citenamefont {G{\'o}mez-Rodr{\'\i}guez}}]{good-cut-and-project-characterization}%
  \BibitemOpen
  \bibfield  {author} {\bibinfo {author} {\bibfnamefont {J.~L.}\ \bibnamefont {Arag{\'o}n}}, \bibinfo {author} {\bibfnamefont {G.~G.}\ \bibnamefont {Naumis}},\ and\ \bibinfo {author} {\bibfnamefont {A.}~\bibnamefont {G{\'o}mez-Rodr{\'\i}guez}},\ }\bibfield  {title} {\bibinfo {title} {Twisted graphene bilayers and quasicrystals: A cut and projection approach},\ }\href@noop {} {\bibfield  {journal} {\bibinfo  {journal} {Crystals}\ }\textbf {\bibinfo {volume} {9}},\ \bibinfo {pages} {519} (\bibinfo {year} {2019})}\BibitemShut {NoStop}%
\bibitem [{\citenamefont {Rodriguez}\ \emph {et~al.}(2008)\citenamefont {Rodriguez}, \citenamefont {McCauley}, \citenamefont {Avniel},\ and\ \citenamefont {Johnson}}]{important-tb-quasicrystal-photonics}%
  \BibitemOpen
  \bibfield  {author} {\bibinfo {author} {\bibfnamefont {A.~W.}\ \bibnamefont {Rodriguez}}, \bibinfo {author} {\bibfnamefont {A.~P.}\ \bibnamefont {McCauley}}, \bibinfo {author} {\bibfnamefont {Y.}~\bibnamefont {Avniel}},\ and\ \bibinfo {author} {\bibfnamefont {S.~G.}\ \bibnamefont {Johnson}},\ }\bibfield  {title} {\bibinfo {title} {Computation and visualization of photonic quasicrystal spectra via bloch’s theorem},\ }\href@noop {} {\bibfield  {journal} {\bibinfo  {journal} {Phys. Rev. B}\ }\textbf {\bibinfo {volume} {77}},\ \bibinfo {pages} {104201} (\bibinfo {year} {2008})}\BibitemShut {NoStop}%
\bibitem [{\citenamefont {Haynes}(2016)}]{history-7}%
  \BibitemOpen
  \bibfield  {author} {\bibinfo {author} {\bibfnamefont {A.}~\bibnamefont {Haynes}},\ }\bibfield  {title} {\bibinfo {title} {Equivalence classes of codimension-one cut-and-project nets},\ }\href@noop {} {\bibfield  {journal} {\bibinfo  {journal} {Ergodic Theory Dynam. Systems}\ }\textbf {\bibinfo {volume} {36}},\ \bibinfo {pages} {816} (\bibinfo {year} {2016})}\BibitemShut {NoStop}%
\bibitem [{\citenamefont {Jiang}\ and\ \citenamefont {Zhang}(2014)}]{soft-matter-numerics-use-for-dft}%
  \BibitemOpen
  \bibfield  {author} {\bibinfo {author} {\bibfnamefont {K.}~\bibnamefont {Jiang}}\ and\ \bibinfo {author} {\bibfnamefont {P.}~\bibnamefont {Zhang}},\ }\bibfield  {title} {\bibinfo {title} {Numerical methods for quasicrystals},\ }\href@noop {} {\bibfield  {journal} {\bibinfo  {journal} {J. Comput. Phys.}\ }\textbf {\bibinfo {volume} {256}},\ \bibinfo {pages} {428} (\bibinfo {year} {2014})}\BibitemShut {NoStop}%
\bibitem [{\citenamefont {Blinov}(2015)}]{paper-preq-important}%
  \BibitemOpen
  \bibfield  {author} {\bibinfo {author} {\bibfnamefont {I.~V.}\ \bibnamefont {Blinov}},\ }\bibfield  {title} {\bibinfo {title} {Periodic almost-schr{\"o}dinger equation for quasicrystals},\ }\href@noop {} {\bibfield  {journal} {\bibinfo  {journal} {Sci. Rep.}\ }\textbf {\bibinfo {volume} {5}},\ \bibinfo {pages} {11492} (\bibinfo {year} {2015})}\BibitemShut {NoStop}%
\bibitem [{\citenamefont {Lu}\ and\ \citenamefont {Birman}(1987)}]{most-important-paper}%
  \BibitemOpen
  \bibfield  {author} {\bibinfo {author} {\bibfnamefont {J.~P.}\ \bibnamefont {Lu}}\ and\ \bibinfo {author} {\bibfnamefont {J.~L.}\ \bibnamefont {Birman}},\ }\bibfield  {title} {\bibinfo {title} {Electronic structure of a quasiperiodic system},\ }\href@noop {} {\bibfield  {journal} {\bibinfo  {journal} {Phys. Rev. B}\ }\textbf {\bibinfo {volume} {36}},\ \bibinfo {pages} {4471} (\bibinfo {year} {1987})}\BibitemShut {NoStop}%
\bibitem [{\citenamefont {Zhang}\ \emph {et~al.}(2022)\citenamefont {Zhang}, \citenamefont {Che}, \citenamefont {Liu}, \citenamefont {Wang}, \citenamefont {Zhao}, \citenamefont {Guan}, \citenamefont {Liu}, \citenamefont {Shi},\ and\ \citenamefont {Zi}}]{history-6}%
  \BibitemOpen
  \bibfield  {author} {\bibinfo {author} {\bibfnamefont {Y.}~\bibnamefont {Zhang}}, \bibinfo {author} {\bibfnamefont {Z.}~\bibnamefont {Che}}, \bibinfo {author} {\bibfnamefont {W.}~\bibnamefont {Liu}}, \bibinfo {author} {\bibfnamefont {J.}~\bibnamefont {Wang}}, \bibinfo {author} {\bibfnamefont {M.}~\bibnamefont {Zhao}}, \bibinfo {author} {\bibfnamefont {F.}~\bibnamefont {Guan}}, \bibinfo {author} {\bibfnamefont {X.}~\bibnamefont {Liu}}, \bibinfo {author} {\bibfnamefont {L.}~\bibnamefont {Shi}},\ and\ \bibinfo {author} {\bibfnamefont {J.}~\bibnamefont {Zi}},\ }\bibfield  {title} {\bibinfo {title} {Unfolded band structures of photonic quasicrystals and moir{\'e} superlattices},\ }\href@noop {} {\bibfield  {journal} {\bibinfo  {journal} {Phys. Rev. B}\ }\textbf {\bibinfo {volume} {105}},\ \bibinfo {pages} {165304} (\bibinfo {year} {2022})}\BibitemShut {NoStop}%
\bibitem [{\citenamefont {Bartolotti}\ and\ \citenamefont {Flurchick}(1996)}]{history-8}%
  \BibitemOpen
  \bibfield  {author} {\bibinfo {author} {\bibfnamefont {L.~J.}\ \bibnamefont {Bartolotti}}\ and\ \bibinfo {author} {\bibfnamefont {K.}~\bibnamefont {Flurchick}},\ }\bibfield  {title} {\bibinfo {title} {An introduction to density functional theory},\ }\href@noop {} {\bibfield  {journal} {\bibinfo  {journal} {Rev. Comput. Chem.}\ ,\ \bibinfo {pages} {187}} (\bibinfo {year} {1996})}\BibitemShut {NoStop}%
\bibitem [{\citenamefont {Davies}\ and\ \citenamefont {Morini}(2024)}]{characterization-1d-fractal-quasicrystal-spectrum}%
  \BibitemOpen
  \bibfield  {author} {\bibinfo {author} {\bibfnamefont {B.}~\bibnamefont {Davies}}\ and\ \bibinfo {author} {\bibfnamefont {L.}~\bibnamefont {Morini}},\ }\bibfield  {title} {\bibinfo {title} {Super band gaps and periodic approximants of generalised fibonacci tilings},\ }\href@noop {} {\bibfield  {journal} {\bibinfo  {journal} {Proc. R. Soc. A}\ }\textbf {\bibinfo {volume} {480}},\ \bibinfo {pages} {20230663} (\bibinfo {year} {2024})}\BibitemShut {NoStop}%
\bibitem [{\citenamefont {Hafner}\ and\ \citenamefont {Kraj{\v{c}}{\'\i}}(1992)}]{QC-brute-force}%
  \BibitemOpen
  \bibfield  {author} {\bibinfo {author} {\bibfnamefont {J.}~\bibnamefont {Hafner}}\ and\ \bibinfo {author} {\bibfnamefont {M.}~\bibnamefont {Kraj{\v{c}}{\'\i}}},\ }\bibfield  {title} {\bibinfo {title} {Electronic structure and stability of quasicrystals: Quasiperiodic dispersion relations and pseudogaps},\ }\href@noop {} {\bibfield  {journal} {\bibinfo  {journal} {Phys. Rev. Lett.}\ }\textbf {\bibinfo {volume} {68}},\ \bibinfo {pages} {2321} (\bibinfo {year} {1992})}\BibitemShut {NoStop}%
\bibitem [{\citenamefont {Ismail-Beigi}\ and\ \citenamefont {Arias}(2000)}]{DFTpp}%
  \BibitemOpen
  \bibfield  {author} {\bibinfo {author} {\bibfnamefont {S.}~\bibnamefont {Ismail-Beigi}}\ and\ \bibinfo {author} {\bibfnamefont {T.}~\bibnamefont {Arias}},\ }\bibfield  {title} {\bibinfo {title} {New algebraic formulation of density functional calculation},\ }\href@noop {} {\bibfield  {journal} {\bibinfo  {journal} {Comput. Phys. Commun.}\ }\textbf {\bibinfo {volume} {128}},\ \bibinfo {pages} {1} (\bibinfo {year} {2000})}\BibitemShut {NoStop}%
\bibitem [{\citenamefont {Jagannathan}(2021)}]{quasicrystal-symm-multifractality}%
  \BibitemOpen
  \bibfield  {author} {\bibinfo {author} {\bibfnamefont {A.}~\bibnamefont {Jagannathan}},\ }\bibfield  {title} {\bibinfo {title} {The fibonacci quasicrystal: Case study of hidden dimensions and multifractality},\ }\href@noop {} {\bibfield  {journal} {\bibinfo  {journal} {Rev. Mod. Phys.}\ }\textbf {\bibinfo {volume} {93}},\ \bibinfo {pages} {045001} (\bibinfo {year} {2021})}\BibitemShut {NoStop}%
\bibitem [{\citenamefont {Bak}(1986)}]{atoms-as-lines}%
  \BibitemOpen
  \bibfield  {author} {\bibinfo {author} {\bibfnamefont {P.}~\bibnamefont {Bak}},\ }\bibfield  {title} {\bibinfo {title} {Icosahedral crystals: Where are the atoms?},\ }\href@noop {} {\bibfield  {journal} {\bibinfo  {journal} {Phys. Rev. Lett.}\ }\textbf {\bibinfo {volume} {56}},\ \bibinfo {pages} {861} (\bibinfo {year} {1986})}\BibitemShut {NoStop}%
\bibitem [{\citenamefont {Kraemer}\ and\ \citenamefont {Sanders}(2013)}]{atoms-as-lines-new}%
  \BibitemOpen
  \bibfield  {author} {\bibinfo {author} {\bibfnamefont {A.~S.}\ \bibnamefont {Kraemer}}\ and\ \bibinfo {author} {\bibfnamefont {D.~P.}\ \bibnamefont {Sanders}},\ }\bibfield  {title} {\bibinfo {title} {Embedding quasicrystals in a periodic cell: dynamics in quasiperiodic structures},\ }\href@noop {} {\bibfield  {journal} {\bibinfo  {journal} {Phys. Rev. Lett.}\ }\textbf {\bibinfo {volume} {111}},\ \bibinfo {pages} {125501} (\bibinfo {year} {2013})}\BibitemShut {NoStop}%
\bibitem [{\citenamefont {Martin}(2020)}]{martin2020electronic}%
  \BibitemOpen
  \bibfield  {author} {\bibinfo {author} {\bibfnamefont {R.~M.}\ \bibnamefont {Martin}},\ }\href@noop {} {\emph {\bibinfo {title} {Electronic Structure: Basic Theory and Practical Methods}}}\ (\bibinfo  {publisher} {Cambridge University Press},\ \bibinfo {year} {2020})\BibitemShut {NoStop}%
\bibitem [{\citenamefont {Bridges}\ and\ \citenamefont {Schuster}(2006)}]{kronecker}%
  \BibitemOpen
  \bibfield  {author} {\bibinfo {author} {\bibfnamefont {D.}~\bibnamefont {Bridges}}\ and\ \bibinfo {author} {\bibfnamefont {P.}~\bibnamefont {Schuster}},\ }\bibfield  {title} {\bibinfo {title} {A simple constructive proof of kronecker's density theorem},\ }\href@noop {} {\bibfield  {journal} {\bibinfo  {journal} {Elem. Math.}\ }\textbf {\bibinfo {volume} {61}},\ \bibinfo {pages} {152} (\bibinfo {year} {2006})}\BibitemShut {NoStop}%
\end{thebibliography}%

\end{document}